\documentclass[12pt]{iopart}

\usepackage{epsfig}
\usepackage{amssymb}
\begin{document}

\title{Azimuthal correlations of forward dijets in d+Au collisions at RHIC}

\author{Cyrille Marquet}

\address{Department of Physics, Columbia University, New York, NY 10027, USA\\
Institut de Physique Th\'eorique, CEA/Saclay, 91191 Gif-sur-Yvette Cedex, France}
\ead{cyrille@phys.columbia.edu}
\begin{abstract}

Measuring correlations between forward dijets in deuteron-gold collisions at RHIC
will further test the Color Glass Condensate (CGC) picture of the nuclear wavefunction
at small$-x,$ which successfully predicted the suppressed production of high-$p_T$
particles at forward rapidities in d+Au collisions: $R_{dA}\!<\!1.$ I present
predictions for the correlation in azimuthal angle between forward dijets in the CGC
framework, with both multiple scatterings and non-linear QCD evolution at small$-x$
taken into account.

\end{abstract}


\section{Forward particle production in p+A type collisions}
Forward particle production in proton-nucleus collisions allows one to investigate the 
non linear QCD dynamics of high-energy nuclei with a probe well understood in QCD. Indeed, while such processes are probing small-momentum (small$-x$) partons in the nuclear wavefunction, only high-momentum partons of the proton contribute to the scattering. For instance, in the case of
two-particle production, the longitudinal fraction of momenta probed in the proton ($x_p$) and in the nucleus ($x_A$) are given by
\begin{equation}
\sqrt{s}\ x_p=|k_1|\ e^{y_1}+|k_2|\ e^{y_2}\ ,\hspace{0.5cm}
\sqrt{s}\ x_A=|k_1|\ e^{-y_1}+|k_2|\ e^{-y_2}
\label{kine}\end{equation}
where $k_1,$ $k_2$ and $y_1,$ $y_2$ are the transverse momenta and rapidities of the final state particles. Therefore, with $\sqrt{s}\gg |k_1|,|k_2|$ and forward rapities $y_1,y_2\!>\!0,$ the process features $x_p\!\lesssim\!1$ and $x_A\!\ll\!1,$ meaning that the scattering involves a well-known dilute hadron $p$ and dense nuclear target $A,$ whose non-linear QCD dynamics can be studied.

The Color Glass Condensate (CGC) framework was quite successful in describing single inclusive particle production at forward rapidities in d+Au collisions at RHIC \cite{jyrev,dhj}. In this work, we focus on forward inclusive two-particle production $pA\!\to\!h_1h_2X$ with $h_1$ and
$h_2$ detected in the proton direction, and in particular on correlations in azimuthal angle between the produced hadrons \cite{klm}. We argue that the second d+Au run at RHIC gives the opportunity to carry out these measurements, and check the relevance of the CGC picture of high-energy nuclei at RHIC energies.

In the following, the $pA\!\to\!h_1h_2X$ cross-section is derived and used in the context of d+Au collisions to predict the azimuthal angle distribution.
 
\section{Forward dijet production}

The kinematic ranges for forward particle measurements at RHIC are such that $x_p\!\sim\!0.5$ and
$x_A\!\sim\!10^{-4}.$ Therefore the dominant partonic subprocess is initiated by valence quarks in the proton and, at lowest order in $\alpha_s,$ the $pA\!\to\!h_1h_2X$ cross-section is obtained from the $qA\to qgX$ cross-section, the quark density in the proton and the appropriate hadron fragmentation functions.

By contrast, the nucleus $A,$ whose partons with small fraction of momentum (mainly gluons) are described by a CGC, cannot be described by a single gluon density. The $qA\!\to\!qgX$ cross section is instead expressed in terms of correlators of Wilson lines (which account for multiple scatterings), with up to a six-point correlator \cite{paper}. Assuming a Gaussian CGC wavefunction allows to express all the correlators in terms of a single function $\Gamma(r,x_A),$ related to the variance of the Gaussian distribution \cite{fgv}. In the large$-N_c$ limit, the non-linear QCD evolution at small$-{x_A}$ is included through the Balitsky-Kovchegov (BK) evolution
\cite{bk-b,bk-k} of $e^{-\Gamma}$.

The cross-section for the production of the quark-gluon dijet (with respective transverse momenta $q_\perp$ and $k_\perp$ and rapidities $y_q\!>\!0$ and $y_k\!>\!0$) then reads \cite{paper}:
\begin{equation}
\fl\frac{d\sigma^{pA\to qgX}}{d^2k_\perp d^2q_\perp dy_k dy_q}\!\propto\!
\frac{\alpha_S C_F}{4\pi^2}x_p q(x_p,\Delta)
\sum_{\lambda\alpha\beta}
\left|I^{\lambda}_{\alpha\beta}(z,k_\perp\!-\!\Delta;{x_A})\!-\!
\psi^{\lambda}_{\alpha\beta}(z,k_\perp\!-\!z\Delta)\right|^2\!F_{x_A}(\Delta)
\label{cs}\end{equation}
with $\Delta=k_\perp+q_\perp$ and $1/z-1=|q_\perp|\ e^{y_q-y_k}/|k_\perp|.$ The different components in (\ref{cs}) are:
\begin{itemize}
\item the quark distribution function in the proton $q(x_p,\Delta)$ (we are working with
$|\Delta|\!\gg\!\Lambda_{QCD}$); in principle the gluon initiated processes $gA\to q\bar{q}X$ and $gA\to ggX$ should also be included, they would contribute for smaller values of $x_p.$ 
\item the Fourier transform of the two-point (dipole) correlator $F_{x_A}(\Delta),$ also called the unintegrated gluon distribution function, given by
\begin{equation}
F_{x_A}(\Delta)=\int\frac{d^2r}{(2\pi)^2}\ e^{-i\Delta\cdot r}
e^{-\Gamma(r,x_A)}\ ;
\label{gdis}\end{equation}
contrary to the dilute proton, the CGC cannot be described only by its gluon distribution
$F_{x_A}(\Delta),$ the $k_T-$factorization formula is not applicable \cite{nszz,bknw} (it would if there was no other $x_A$ dependence in (\ref{cs}), as in single gluon production \cite{kt,gprod}).
\item the $k_T-$factorization breaking terms (whose simple form
is due the use of a Gaussian CGC wavefunction) $\displaystyle\sum_{\lambda\alpha\beta}
\left|I^{\lambda}_{\alpha\beta}(x_A)-\psi^{\lambda}_{\alpha\beta}\right|^2$ with
\begin{equation}
I^{\lambda}_{\alpha\beta}(z,k_\perp;{x_A})
=\int d^2q_\perp \psi^{\lambda}_{\alpha\beta}(z,q_\perp) F_{x_A}(k_\perp\!-\!q_\perp)\ ;
\label{split}\end{equation}
as the valence quark emits the virtual gluon (the associated wavefunction is $\psi^{\lambda}_{\alpha\beta}$ where $\lambda,$ $\alpha$ and $\beta$ are polarization and spin indices) it interacts coherently with the dense small$-x$ gluons in $A,$ which modifies the $q\!\to\!qg$ splitting.
\end{itemize}
In practice, the initial condition $\Gamma(r,x_0)=r^2Q_{s_0}^2\ln[e+1/(r^2\Lambda_{QCD}^2)]/4$
is used with the choice $x_0=0.01$ and the initial saturation scale
$2\pi\ Q_{s_0}^2=2\ \mbox{GeV}^2.$ Then $e^{-\Gamma}$ is evolved with the BK equation, allowing to compute (\ref{gdis}), (\ref{split}) and the cross-section (\ref{cs}).

\section{Dijet correlations in azimuthal angle}

We will now use the inclusive two-particle spectrum (\ref{cs}) to investigate the process
$dAu\!\to\!h_1h_2X,$ with $\sqrt{s}\!=\!200\ \mbox{GeV}.$ In particular we shall study 
the $\Delta\phi$ spectrum where $\Delta\phi\!=\!\phi_1\!-\!\phi_2$ is the difference between the azimuthal angles of the measured particles $h_1$ and $h_2.$ We will study the normalized 
$\Delta\phi$ distribution
\begin{equation}
\frac{1}{\sigma}\frac{d\sigma}{d\Delta\phi}\equiv
\left(\frac{d\sigma^{h{\cal T}\to h_1h_2X}}{dp_{T_1}dp_{T_2}dy_1 dy_2}\right)^{-1}
\frac{d\sigma^{h{\cal T}\to h_1h_2X}}{dp_{T_1}dp_{T_2}dy_1 dy_2 d\Delta\phi}
\label{obs}\end{equation}
where $k_1\!=\!(p_{T_1},\phi_1)$ and $k_2\!=\!(p_{T_2},\phi_2)$ are the transverse momenta of the measured hadrons and $y_1$ and $y_2$ are their rapidities. We take into account both situations where either the quark or the gluon fragments into the most forward hadron. Our results do not include convolution with fragmentation functions which should be done eventually, however we except this has little impact on the observable we are considering here; for $p_T-$spectra this would be not be case.

The main features of the $\Delta\phi$ spectrum (\ref{obs}) are the following. First we obtain that the perturbative back-to-back peak of the azimuthal angle distribution is recovered for very large transverse momenta. Then, this back-to-back correlation is reduced by the initial state saturation effects included in our CGC approach: as the transverse momenta decrease closer to the saturation scale ($Q_s\!\simeq\!2\ \mbox{GeV}$), the angular distribution broadens. Finally, we notice that at RHIC energies, this does not lead to a complete disappearance of the back-to-back peak.

As can be seen from the kinematics (\ref{kine}), the most forward of the two particles essentially determines the value of $x_h$ while the most central one determines the value of $x_A.$ In order to quantitatively study the effect of the CGC evolution, the ideal situation would be to keep
$x_h$ fixed and to vary $x_A.$ In practice, this is better realized by fixing the rapidity and momentum of the most forward particle and by varying the kinematics of the other. Note that doing the opposite would emphasize the $x_h$ evolution of $q(x_h,\Delta),$ rather than focus on the
$x_A$ evolution of $F_{x_A}.$ Moreover, the cross-section (\ref{cs}) is quite sensitive to choice of factorization scale in the quark density, so it is better to keep $x_h$ constant. Note that varying the rapidities at fixed $y_1\!-\!y_2$ would keep the product $x_h x_A$ constant, and would force a competition between the evolution of $q(x_h,\Delta)$ with increasing $x_h$ and the CGC evolution with decreasing $x_A.$

In Figure 1a, we have studied the $\Delta\phi$ spectrum (\ref{obs}) in the situation in which
$p_{T_1}\!=\!3.5\ \mbox{GeV},$ $p_{T_2}\!=\!2\ \mbox{GeV},$ $y_1\!=\!3.5$ and $y_2$ is varied from $1.5$ to $2.5.$ As $y_2$ increases, the value of $x_A$ decreases and the suppression of the azimuthal correlation is more important. However the effect is quite small, because the increase of the saturation scale with decreasing $x_A$ is rather slow. In Figure 1b, we investigate the situation for which $p_{T_1}\!=\!5\ \mbox{GeV},$ $y_1\!=\!3.5,$ $y_2\!=\!2$ and
$p_{T_2}$ is varied $1.5\ \mbox{GeV}$ to $3\ \mbox{GeV}.$ As $p_{T_2}$ decreases, it gets closer to the saturation scale $Q_s$ (which also slightly increases as $x_A$ decreases), and the suppression of the azimuthal correlation increases. Varying $p_{T_2}$ at fixed $y_2$ allows to probe the ratio $p_{T_2}/Q_s$ over a larger range, so the effect is much bigger than when varying $y_2$ at fixed $p_{T_2}.$

\begin{figure}[t]
\begin{center}
\epsfig{file=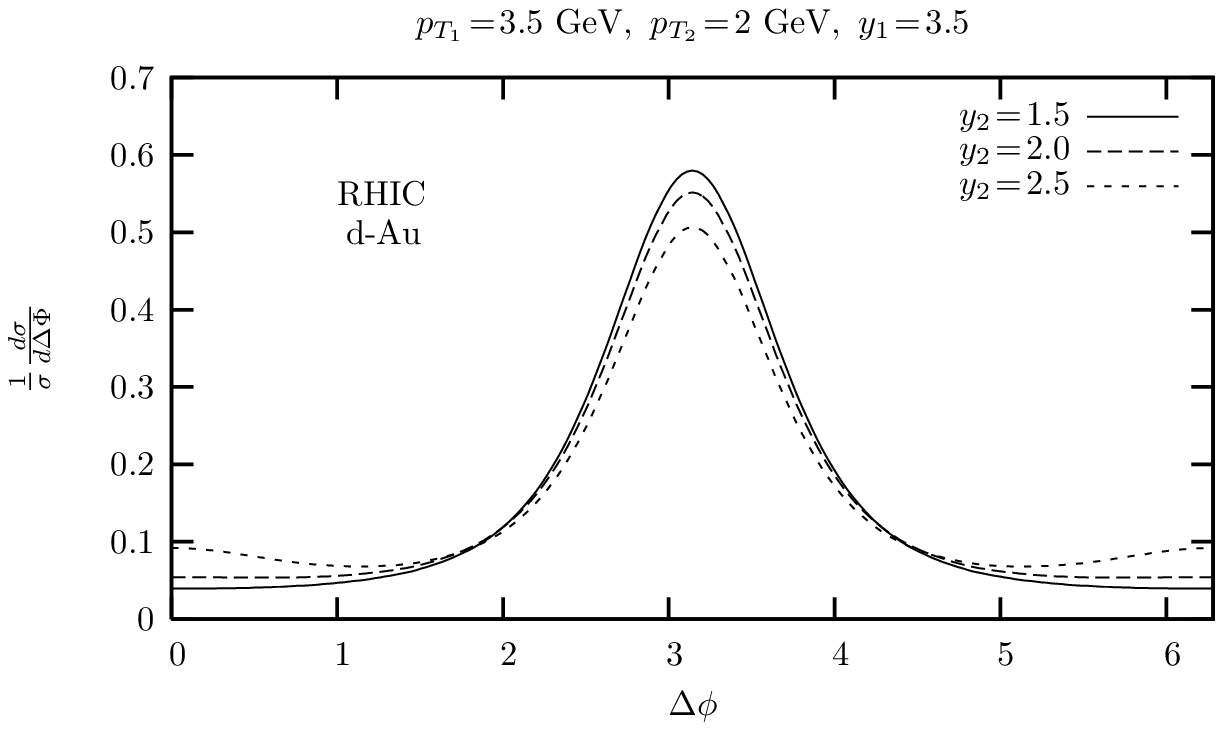,width=7.4cm}
\hfill
\epsfig{file=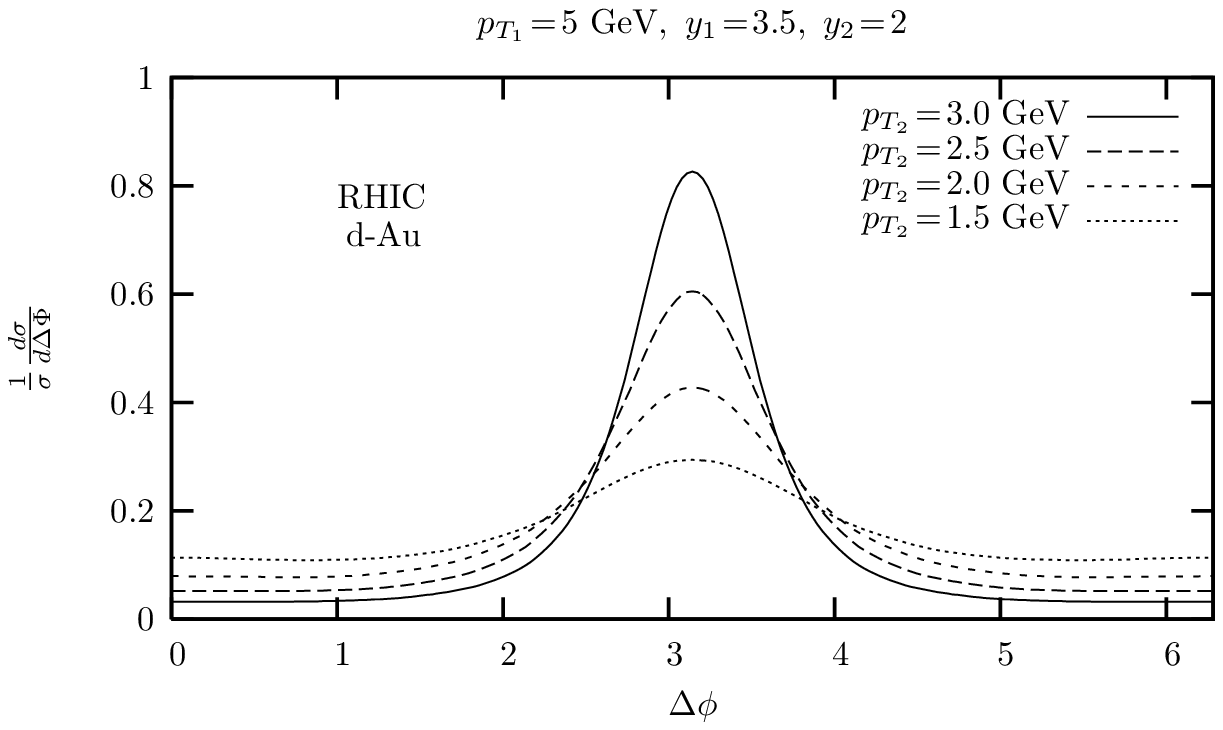,width=7.4cm}
\caption{The $\Delta\phi$ spectrum (\ref{obs}) in two situations with the RHIC energy
$\sqrt{s}\!=\!200\ \mbox{GeV/nucleon}.$ In both cases, by increasing $y_2$ or decreasing $p_{T_2},$ the correlation in azimuthal angle is suppressed as the value of $x_A$ probed in the process decreases. Varying $p_{T_2}$ at fixed $y_2$ is much more efficient as the ratio $p_{T_2}/Q_s$ covers a larger range.}
\end{center}
\end{figure}

Experimental measurements of two-particle correlations in azimuthal angle have been performed in
d+Au collisions at RHIC by the PHENIX and STAR collaborations. The measurements of STAR
\cite{star} with $\pi^0$ at forward rapidity and charged hadrons at mid rapidity are qualitatively consistent with a suppression of the back-to-back peak with respect to p+p collisions. By contrast, the measurements of PHENIX \cite{phenix} do not show any evidence of a suppression of the back-to-back peak, but they probe values of $x_A$ which are bigger than $0.01.$ It may very well be that the CGC picture breaks down for values of $x_A$ bigger than $0.01,$ and it justifies our choice not to start the small$-x_A$ evolution at a higher value.

Future measurements of azimuthal correlations between forward particles in d+Au collisions, which could be carried out at RHIC, will allow quantitative tests of the CGC. Our predictions for the fully differential cross section (\ref{cs}) are not directly comparable with data: once the cuts used by the experiments become available, integrations over the kinematic variables should be performed. Measurements in p+Pb collisions at the LHC would reach $x_A\!\sim\!10^{-5},$ and could test even better the QCD evolution at small$-x.$

\end{document}